# MULTIMEDIA DATABASE APPLICATIONS: ISSUES AND CONCERNS FOR CLASSROOM TEACHING

Chien Yu and Teri Brandenburg

Department of Instructional Systems & Workforce Development,
Mississippi State University
cyu@colled.msstate.edu; tbrandenburg@colled.msstate.edu

## ABSTRACT

*The abundance of multimedia data and information is challenging educators to effectively search, browse, access, use, and store the data for their classroom teaching. However, many educators could still be accustomed to teaching or searching for information using conventional methods, but often the conventional methods may not function well with multimedia data. Educators need to efficiently interact and manage a variety of digital media files too. The purpose of this study is to review current multimedia database applications in teaching and learning, and further discuss some of the issues or concerns that educators may have while incorporating multimedia data into their classrooms. Some strategies and recommendations are also provided in order for educators to be able to use multimedia data more effectively in their teaching environments.*

## KEYWORDS

*Multimedia Database, Multimedia, Classroom Teaching,*

## 1. INTRODUCTION

For the past few decades, there has been a significant proliferation of digital media. The growth of the Internet and the development of streaming-video technologies have increased educators' ability to execute teaching strategies to meet a multiplicity of student learning styles [1]. However, the huge amount of multimedia data and information could also challenge educators to efficiently search, browse, access, use, and store the data for their classroom teaching.

Digital media benefit students' learning in many different ways. The developments in computer technology and digital video along with broadband technologies have offered students new and improved learning opportunities. The high-speed Internet also enables individuals to use multimedia representations of enriched audio and video to create complex social networks in which people communicate, exchange ideas, explore, and learn [2]. As a result, educational uses of multimedia cannot go unnoticed in today's classrooms. Digital media provide excellent opportunities for creating engaging learning environments, and help students acquire creative, communicative, and collaborative skills. New media, including Web 2.0 applications such as social networking, blogs, wikis, YouTube, and podcasting have become pervasive and are becoming an integral component of learning environments. The growing popularity of these new media suggests that educators will need to understand these new possibilities of emerging technologies and incorporate them into the teaching and learning environment.

                                                                            1

However, today's multimedia databases contain much more than just text and numbers--they include images, sounds, video, or other scanned documents, and/or data. Many educators may still be found teaching or searching for their information with conventional methods, but often the conventional methods do not function well with multimedia data. Educators need to effectively interact and manage a variety of digital media files, too. The purpose of this study is to review the current applications of multimedia databases in teaching and learning, and further discuss some of the issues or concerns that educators may have while incorporating multimedia data to their classrooms. This study is not meant to discuss the design or technical issues of multimedia databases (e.g., indexing, querying, or retrieving data), rather to discuss from practical perspectives related to the applications, concerns or issues that educators may have while incorporating multimedia data into their teaching, and to further discuss some tips and strategies that educators can use to search, access, and store multimedia data more effectively for their classroom teaching.

## 2. WHAT IS A MULTIMEDIA DATABASE?

### 2.1. Definitions: Multimedia vs. Multimedia Database

With an increased availability of digital information options today, a commonly accepted definition of *multimedia* is a combination of different media (i.e., text, pictures, sounds, video, animations, etc.) used to present multimodal information in conjunction with computer technology [1]. Due to the growing delivery of media by computer and the merging of increasingly powerful computer-based authoring tools with Internet connectivity, it seems that the term "multimedia" is now firmly associated with computer-based delivery. Although the term has not always been associated with computers [3], Gonzalez et al. [4] asserted that "multimedia cannot be experienced without the technology because it is the technology that creates the experience" (¶ 9).

A *multimedia database* is a collection of related multimedia data. Common multimedia data types that can be found in a multimedia database include the following:

- Text
- Graphics: drawing, sketches, and illustrations
- Images: color and black & white pictures, photographs, maps and paintings
- Animation sequences: animated images or graphic objects
- Video: a sequence of images (frames), typically recording a real-life event and usually produced by a video recorder
- Audio: generated from an aural recording device
- Composite multimedia: a combination of two or more of the above data types [5]

Among various media types, media can be divided into two major classes: Continuous and discrete [6]. Media that are changing with time such as audio and video are called continuous media; and time-independent media such as text, still images, and graphics belong to discrete media [6].

## 2.2. Understanding Multimedia Database Characteristics

A multimedia database contains various data types such as text, images, graphic objects (including pictures, drawings and illustrations), animation sequences, video and audio. Additionally, Kalipsiz [6] summarized some characteristics of multimedia data as below:

- *Lack of structure*: Multimedia data often are not quite structured; therefore, standard indexing and/content-based search and retrieval may not be available.

- *Temporality*: Different multimedia data types have different requirements. For example, some multimedia data types such as video, audio, and animation sequences have temporal requirements that have implications on their storage, manipulation and presentation, but images, video and graphics data have spatial constraints in terms of their content.

- *Massive Volume*: Usually, the data size of multimedia is large such as video; therefore, multimedia data often require a large storage device.

- *Logistics*: Non-standard media can complicate processing. For example, a multimedia database application requires using compression algorithms.

## 2.3. Multimedia Database Applications

Multimedia database applications are different from the traditional database applications in the structure of the multimedia objects and media where the multimedia objects are stored and retrieved [7].

Multimedia data is diverse with different characteristics. Because of the audio-visual nature of multimedia data types, they are usually complex data composed of other data that can also be complex. Aygun, et al. [7] indicated that multimedia objects are multidimensional and hierarchically structured, and can have some relations among them. For example, image data does not have temporal behavior because there is no time associated with it, while video has both temporal and spatial behavior because the image sequences of the video should be displayed in order and in some dedicated time [7].

## 3. APPLICATIONS OF MULTIMEDIA DATABASES FOR CLASSROOM TEACHING

Knowing how to best use and apply multimedia database for classroom teaching can be challenging for teachers. However, to use a multimedia database, teachers do not have to be multimedia experts. Although multimedia data are diverse and complex in their structure, teachers can still apply multimedia data from databases to facilitate their classroom teaching, whether in traditional face-to-face or distance education courses.

### 3.1. Face-To-Face Courses

Multimedia databases are capable of reinforcing the traditional face-to-face classroom. The use of multimedia data can foster and develop cognitive engagement through the ability to attract and hold students' attention and focus [1]. For example, Microsoft Word's grammar checker is a multimedia database application [8]. If a student makes a spelling or grammar error, the system will show the error and can further correct the mistake.

Multimedia databases can not only be used to help students integrate interrelated content areas, but also make learning more meaningful. For example, sound or auditory applications in teaching may actively engage students in analyzing, synthesizing and evaluating information and constructing knowledge [9]. Using multimedia databases, students can ask questions, interpret information, find answers, then further foster their critical thinking and construct their knowledge.

The World Wide Web has enabled access to numerous multimedia databases. The Internet itself is the largest multimedia database. Abundant Internet resources provide teachers with multimedia databases as cognitive tools and enable teachers to integrate multimedia data into their instruction.

### 3.2. Distance Education/E-Learning Courses

With the growing amount of users on the Internet, distance education courses have become more and more available and popular [8]. Distance education is commonly defined as a formal learning activity that occurs when students and instructors are separated by geographical distance or by time and is often supported by technology such as television, videotape, computers, Internet or mail. In order to facilitate the teaching and learning environment, web browsers, audio/video communication tools and data conferencing tools are widely used also [10].

Multimedia databases can not only create enormous opportunities for improving the instructional process, but also help shape teachers' consciousness and perception in using information to enhance the learning environment. For example, Wang and Lin [8] developed a novel English distance learning system through multimedia database and Internet technologies to store English articles, dialogs, videos, and also mistakes that students make. Because of the database, teachers can understand the most frequent mistakes, and students can improve and practice their English abilities in a realistic learning environment through several distinct approaches.

## 4. ISSUES AND CONCERNS OF USING MULTIMEDIA DATA IN CLASSROOM TEACHING

Because technology continues to change dramatically, the different data types may require special methods for optimal storage, access, indexing, and retrieval [6]. Information retrieval is dependent upon the design and implementation of systems.

### 4.1. Data Availability

Candan, Lemar & Subrahmanian [11] stated that multimedia objects have temporal and spatial aspects that do not exist in more traditional data objects; therefore, visualization of the results of a multimedia query requires specification of the visualization parameters along with the query. Some of the multimedia databases are searchable with search engines on their own website; some offer data files that teachers may download to their own computers. As a result, it may take time and expertise for teachers to search and locate relevant and useful databases.

## 4.2. File Format and Size

As mentioned, multimedia data is comprised of text, images, graphics, video, audio, etc. There is an overwhelming number of file representations for these different types of data, including TIFF, BMP, PPT, IVUE, FPX, JPEG, MPEG, AVI, MID, WAV, DOC, GIF, EPS, PNG, etc. Because of restrictions on the conversion from one format to the other, the use of the data in a specific format has been limited as well [12].

In addition to the fact that multimedia objects are complex in their file formats, they are large too. For example, each of the following takes 1 Megabyte of storage in uncompressed form: six seconds of CD-quality audio, a single 640x480 color image with 24 bits/pixel, a single frame of (1/30 second) CIF video, or one digital X-ray image (1024x1024) with 8 bits/pixel [13].

## 4.3. Data Storage and Retrieval

A multimedia database system includes a multimedia database management system and a multimedia database. The database management system manages the multimedia database; a multimedia database is the multimedia data being managed [14]. Why is a database needed? It is for storing and retrieving data more efficiently.

Multimedia information (e.g., text, graphics, audio, video, etc.) has to be managed differently depending on the type of data. However, efficient retrieval of data depends on the database system. Inefficiencies of traditional retrieval approaches could result in a demand for teachers to understand techniques that can manipulate the multimedia data.

## 4.4. Search Engines

Current multimedia databases play an important role on the Web. As Johnson [12] indicated, "There is the desire for Internet multimedia search engines capable of searching and locating the relevant sources containing the desired media types given a description of the specific content" (p. 1/4). Therefore, teachers will want to know how to search and manage multimedia data on both the Internet and intranet, and how to keep up with the explosive increase in multimedia databases.

## 4.5. Teachers' Technology Skills and Attitude

As technology develops, the use of technology in a classroom has become a main concern. However, many teachers would have a hard time consolidating their skills in the use of the existing technologies for classroom instruction, and the high uncertainty of emerging technologies makes it even more difficult for teachers to develop the level of expertise needed to incorporate technology into the classroom. As a result, students' classroom practice may not meet student expectation, especially in the area of integration and use of multimedia [15] because today's students are often far more skilled at using digital media than most of their teachers. Torrisi-Steele [16] stated, "The effective integration of multimedia in the curriculum depends not on the technology itself but rather on educators' knowledge, assumptions, and perceptions" (p. 26).

# 5. STRATEGIES FOR USING MULTIMEDIA DATA IN CLASSROOM TEACHING

As Web applications grow, the need for efficient multimedia databases will become essential. Knowing how to effectively access various databases will become increasingly important for teachers as well.

## 5.1. Pedagogical Concerns

Today's technologies make possible the use of multimedia by helping to move learning beyond a primarily text-based and linear arena into the cyclical world of sights, sounds, creativity, and interactivity. But, the challenge is whether the essence of multimedia can be integrated into an essential discipline [4]. If some pedagogical design principles can be used effectively, multimedia can permit greater individualization, in turn fostering improved learning, learner satisfaction, and retention rates [17].

With the large amount of information in databases, it is necessary for teachers to guide students through meaningful learning activities so that they can learn how to use databases to facilitate their own thinking [9]. Jonassen [18] suggests some strategies for classroom applications, for example, teaching with databases should proceed gradually, starting from having students work with completed databases, to partially completed databases and then to databases created by students themselves. This process suggests how to provide different scaffolding to students with different database skills. Students' critical thinking and problem-solving skills should be developed gradually from learning with guidance [9].

## 5.2. Data Searching

As more multimedia information becomes available, the need for efficient browsing, searching, and retrieving of information increases. Petkovic & Jonker [19] described the increasing demand to manipulate the data based on the content. Furthermore, Johnson [18] indicated that three approaches can be used to represent the content of multimedia data. They are: keyword-based, feature-based, and concept-based approaches.

- *Keyword-based approach*: The multimedia content is described by the user through annotations.

- *Feature-based approach*: A set of features can be used for representing and retrieving the multimedia data. Many multimedia databases allow users to query a database by specifying keywords and/or image characteristics. For example: general information like color, texture, shape, speed, position, or particular applications like fingerprint recognition or medical images.

- *Concept-based approach*: Concepts are subsequently used to interpret the content of the data and to retrieve the data itself. "This is usually an application domain specific process and occasionally requires the intervention of the user" [18] (p. 1/3).

For efficient browsing and searching, querying the multimedia database is an important part of a database system. According to Kalipsiz [6], multimedia data queries can be divided into four different types. They are: a) Keyword query, b) Semantic query, c) Visual query, and d) Video query. Keyword querying only uses well-defined queries, while semantic and visual querying are designed to use the fuzzy (more approximate, less precise) query method [6].

Querying in a multimedia database is quite different from querying in standard alphanumeric databases. If querying conventional databases, which consist of text or numerical data, a query is often represented in the form of text or a numerical value [20]. Besides the fact that browsing takes on added importance in a multimedia environment, queries can contain multimedia objects input by the user; the results of these queries are based not on perfect matches but on degrees of similarity [21].

While a content-based approach is preferable in a multimedia database system [22], Yoshitaka & Ichikawa [20] further indicated a query-by-example approach for content-based representation, which is a method of query specification and allows a user to specify a query by giving examples. For example, an image object is retrieved by the shape of objects, by specifying colors and their spatial distribution in the image, or by a specific pattern appearing in an image.

## 6. CONCLUSION

Computer technology and Internet resources provide teachers with both challenges and opportunities in teaching with multimedia databases. Since teachers are the key to providing quality learning experiences for students, they need to make the best use of multimedia information within the framework of educational theories and learning principles. The growing popularity of multimedia sources suggests that educators will need to conceptualize these new possibilities of emerging technologies and incorporate them within a concrete framework of teaching and learning. For effective teaching and learning, teachers are also required to search and present the best selection of multimedia resources.

However, multimedia information and databases are quite different from standard traditional alphanumeric databases; therefore, this study provided a brief introduction to the effects of multimedia databases in classroom teaching and highlighted some of the concerns or problems that teachers may have faced in their instruction utilizing multimedia databases. Multimedia information and databases will be valuable for teachers who want to integrate multimedia resources into curriculum and teach with databases as cognitive tools. With effective application of multimedia information and databases, teachers will be better able to achieve successful instructional outcomes for various content areas.

**Authors**

Dr. Chien Yu is currently serving an Associate Professor at the Department of Instructional Systems and Workforce Development of Mississippi State University. Her research agenda include distance education, technology integration, educational leadership, instructional design and media development. She can be reached at cyu@colled.msstate.edu.

Dr. Teri Brandenburg is an Assistant Professor at the Department of Instructional Systems and Workforce Development of Mississippi State University. She can be reached at tbrandenburg@colled.msstate.edu.